\documentclass[12pt]{article}
\usepackage[a4paper,margin=1in]{geometry}
\usepackage{amsmath,amssymb,bm}
\usepackage{siunitx}
\usepackage{booktabs}
\usepackage{graphicx}
\usepackage{enumitem}
\usepackage[superscript,biblabel]{cite}
\usepackage{caption,subcaption}
\usepackage{float}
\usepackage{tikz}
\usetikzlibrary{decorations.pathmorphing}
\usepackage{hyperref}
\usepackage{setspace}
\title {Living Helices in Fluctuating Polymer Chains: Cooperative Nucleation, Dynamics, and Lifetime}

\author{Biman Bagchi*\\
Solid State and Structural Chemistry Unit\\
Indian Institute of Science, Bengaluru 560012, India\\
\texttt{profbiman@gmail.com}}

\date{}

\begin{document}
\maketitle

\begin{abstract}

Helical segments in polymer chains are often transient, finite, and dynamically evolving, yet their origin and stability remain incompletely understood. Here we develop a minimal coarse-grained statistical-mechanical theory that explains how such “living helices” emerge in fluctuating polymer systems.
Using a three-state model with cooperative interactions, we show that helix formation proceeds through a multistep nucleation mechanism. An initial constrained pre-nucleus forms first, followed by cooperative stabilization that promotes the growth of finite helical segments. The resulting free-energy landscape naturally favors marginally stable helices whose size is determined by a competition between cooperative gains and nonlinear penalties arising from stiffness, torsional strain, and solvent fluctuations.
By formulating the dynamics as a stochastic process in segment size, we derive analytical expressions for both formation times and lifetimes within a mean first-passage framework. For representative parameters relevant to flexible polymers and peptide segments, the theory predicts characteristic timescales in the nanosecond to sub-microsecond range.
The present analysis supports a view of living helices as finite, mobile excitations whose
stability is controlled by cooperativity, boundary motion, and solvent-induced
fluctuations.

\end{abstract}

\section{Introduction}
\noindent

The formation of helical structures in polymer chains is a subtle and
fundamentally non-generic phenomenon. In flexible polymers, attractive
interactions usually favor collapse into disordered globular states,
whereas the emergence of ordered helices requires additional physical
ingredients such as bending stiffness, torsional constraints, and
directional interactions including hydrogen bonding. Recent theoretical
and simulation studies have shown that stiffness can stabilize a variety
of ordered morphologies, including rods, toroids, and helices, depending
on the competition between elasticity and attraction. In this context,
helices represent particularly delicate structures that arise only when
geometric, energetic, and cooperative constraints are simultaneously
satisfied.

Single-chain helices are ubiquitous in both biological and synthetic
macromolecules. The best-known example is the $\alpha$-helix in proteins,
stabilized by intramolecular hydrogen bonding, as first elucidated by
Pauling and Corey.\cite{Pauling1951} Helical motifs are central to the
structure and function of many proteins and biopolymers, including
keratin, myoglobin, amylose, and RNA secondary structures.\cite{BrandenTooze,ImbertyPerez,TinocoBustamante}
In synthetic systems, helices occur in polyisocyanates, synthetic
polypeptides such as poly($\gamma$-benzyl-L-glutamate), substituted
polyacetylenes, methacrylate-based polymers, and more recently in
foldamers and peptoid systems.\cite{GreenEtAl,DotyYang,NakaiFujiki,GellmanFoldamers}
These examples underscore the broad ubiquity of helical organization,
while simultaneously raising a fundamental question: under what physical
conditions does a polymer chain select a helical conformation over
competing disordered or collapsed states?

From a broader perspective, helix formation may also be viewed within
the energy-landscape framework of protein folding developed by Dill,
Onuchic, Wolynes, and co-workers.\cite{Dill1995,Onuchic1997,Dill1997,OnuchicWolynes2004}
Within this picture, secondary structures such as $\alpha$-helices and
$\beta$-hairpins emerge as local free-energy minima along folding
pathways and often form on nanosecond to submicrosecond timescales.
Most such approaches, however, focus on global folding trajectories and
equilibrium structures, whereas the present work addresses finite
helical segments as cooperative stochastic excitations in a fluctuating
polymer chain.

Classical theories of helix--coil transition, notably those of Zimm and
Bragg and Gibbs and DiMarzio,\cite{ZimmBragg1958,GibbsDiMarzio1958}
provided a successful equilibrium description of the competition between
energetic stabilization and configurational entropy loss associated with
helical ordering. These approaches emphasize the cooperative nature of
helix formation, wherein stabilization requires the buildup of multiple
favorable contacts along the chain.

More recently, we developed a microscopic theoretical framework for
coil-to-helix transitions in generic polymer systems, with emphasis on
the minimal physical ingredients required for helix stabilization.\cite{BagchiMorphology, Bagchi2026}
Two complementary mechanisms were identified. The first is a
packing-driven route, in which steric constraints and excluded-volume
effects select an optimal helical geometry characterized by a preferred
radius and pitch. The second is an interaction-driven route based on
directional or periodic sticky interactions, such as hydrogen bonding,
where commensurability between interaction spacing and backbone geometry
promotes cooperative helix formation. While the packing model leads to
spontaneous symmetry breaking between left- and right-handed helices,
the interaction-driven model connects naturally with classical
helix--coil theories through cooperative stabilization. These results
demonstrated that similar helical structures may emerge from distinct
microscopic origins with fundamentally different kinetic pathways and
stability criteria.

Despite extensive work on equilibrium structures and thermodynamics, a
comprehensive theory for the \emph{dynamics} of coil-to-helix transitions
remains lacking. This gap is particularly important because experiments
and simulations increasingly indicate that helices are intrinsically
finite, heterogeneous, and dynamically generated objects rather than
infinitely extended periodic structures. AFM studies on synthetic
polymers such as polyacetylenes and polyisocyanates reveal finite
helical segments with well-defined pitch and handedness,\cite{FujikiAFM,ZhangAFM}
while cryo-EM studies on biological filaments and supramolecular
assemblies show local helical order persisting even in the absence of
global periodicity.\cite{EgelmanCryoEM} Spectroscopic probes including
circular dichroism and NMR further support a picture of dynamically
interconverting helical and coil segments.\cite{GreenCD,DysonWright}

These observations suggest that helix formation is not governed by a
single universal mechanism but may arise from multiple competing and
nearly degenerate pathways involving packing, hydrogen bonding, and
local cooperative interactions. Such a situation naturally leads to
\emph{accidental degeneracy}, in which structurally similar helical
states emerge from distinct microscopic mechanisms with comparable free
energies but different kinetic pathways. Understanding how this
degeneracy influences helix nucleation, growth, migration, and decay is
therefore essential for a unified dynamical theory of coil-to-helix
transitions.

Several important issues nevertheless remain unresolved. Most existing
theories focus primarily on equilibrium properties and do not explicitly
describe the stochastic dynamics of helix nucleation and decay.
Furthermore, the interplay among stiffness, cooperativity, and solvent
effects is often incorporated only implicitly. Experimental and
simulation studies also indicate that helices are finite, fluctuating,
and heterogeneous objects rather than perfectly periodic structures,
requiring a framework that treats them as dynamical excitations.

In this work we develop such a framework and address two closely related
physical issues: accidental degeneracy and the emergence of finite
``living helices.'' Here we do not address the selection of helical handedness. We focus instead on the formation, 
lifetime, and migration of finite helical segments along a fluctuating polymer chain. 

\subsection*{(i) Accidental degeneracy and spontaneous chirality selection}

In the absence of intrinsic chiral bias, the Hamiltonian of a polymer
with isotropic interactions and symmetric local constraints is invariant
under inversion of handedness. Left- and right-handed helices are
therefore energetically degenerate, and the selection of handedness
emerges dynamically through fluctuation-induced symmetry breaking. The
chosen handedness is subsequently propagated along the chain through
cooperative interactions. A theoretical description of this process
requires going beyond equilibrium statistical mechanics to include
stochastic dynamics and fluctuation-induced nucleation.

\subsection*{(ii) Living helices and marginal stability}

A central feature of polymer helices is their intrinsic marginal
stability. Short helical segments are unstable because local interactions
alone are insufficient to stabilize them, whereas longer segments gain
stability through cooperative buildup of favorable interactions such as
hydrogen bonding. In the regime where the free-energy difference between
helix and coil states is small, the system enters a ``living helix''
state in which finite helical segments continuously form, dissolve, and
migrate along the chain.

The motion of these segments arises from successive formation and
melting events at their boundaries, producing an effective stochastic
propagation along the polymer contour. A quantitative description
therefore requires an explicit treatment of nucleation kinetics,
domain-wall motion, and the associated free-energy landscape.

\subsection*{(iii) Role of water fluctuations and local stabilization of the open state}

An important physical ingredient that has not been explicitly incorporated
in existing helix--coil theories is the role of solvent fluctuations,
particularly those of water. In systems such as polypeptides and nucleic
acids, water molecules can form hydrogen bonds with exposed backbone or
base sites, thereby stabilizing locally open (coil) configurations.

This leads to a situation in which the stability of the helical state is
not governed by a uniform global free-energy difference, but instead by a
spatially varying local environment. The effective free-energy landscape
along the chain is therefore heterogeneous, with regions that favor
helical order interspersed with regions that favor the open state.

From a statistical-mechanical perspective, this situation is analogous to
a random-field Ising system, in which local fields bias the state of each
site. In the present context, these local fields arise from solvent
fluctuations and hydration dynamics, and may be either quenched or
slowly varying. Their presence has profound consequences for the dynamics
of helix formation, leading to pinning of helical segments, nucleation of
bubbles, and broad distributions of lifetimes and domain sizes.


\noindent
In this work, we develop a coarse-grained free-energy theory for the formation, stability, 
and dynamics of finite helical segments in polymer chains with cooperative interactions. 
Our approach explicitly incorporates (i) local helix--coil energetics, (ii) cooperative 
stabilization, (iii) stiffness-induced penalties such as bending and torsional constraints, 
and (iv) solvent-induced heterogeneity. Within this framework, helices emerge as finite, 
marginally stable structures whose properties are governed by a competition between 
collective stabilization and nonlinear penalties.

We formulate the dynamics of helix formation and decay as stochastic processes in segment 
size and analyze them using a mean first-passage-time framework. This allows us to 
distinguish between formation and lifetime as fundamentally different processes. In 
addition, we show that chirality arises dynamically through spontaneous symmetry breaking 
during a multistep nucleation process, providing a unified picture of helix formation as a 
cooperative, stochastic phenomenon.

We construct a free-energy functional that incorporates:
(i) local helix--coil energetics,
(ii) cooperative stabilization due to nearest-neighbor interactions,
(iii) torsional and bending penalties associated with helical geometry, due to stiffness
and
(iv) spatially varying local fields representing solvent-induced
stabilization of the open state.

Within this framework, we analyze the stochastic dynamics of helix
formation and melting, and derive analytical expressions for key
quantities such as the lifetime of a helical segment of length $n$ and
the effective propagation of helical domains along the chain.

The remainder of the paper is organized as follows. In Sec.~2, we introduce
the coarse-grained three-state model and the corresponding free-energy
functional incorporating cooperativity, chirality, and solvent effects.
Sec.~3 outlines the thermodynamic framework. In Sec.~4, we formulate the
stochastic dynamics and master equation consistent with the Hamiltonian.
Sec.~5 develops the microscopic picture of local bistability and
cooperative nucleation. In Sec.~6, we derive a baseline kinetic theory
for the lifetime of helical segments within a linear free-energy model.
Sec.~7 extends the analysis to nonlinear free-energy landscapes, leading
to finite, marginally stable helices and activated dynamics. In Sec.~8,
we analyze migration of helical segments via stochastic boundary motion.
Sec.~9 discusses the connection to DNA bubble dynamics and solvent-induced
heterogeneity. Sec.~10 presents a kinetic theory of chirality selection
and multistep nucleation. Numerical estimates of lifetimes and sizes are
given in Sec.~11, followed by a mean first-passage formulation of helix
formation rates in Sec.~12. We conclude with a discussion of the
implications and possible extensions of the present work.

\begin{figure}[t]
\centering
\begin{tikzpicture}[scale=1.2]

\node at (0,2.0) {\small (a) Polymer chain};

\foreach \x in {0,0.8,1.6,2.4,3.2}
{
    \fill[black] (\x,0) circle (0.08);
}

\foreach \x in {0,0.8,1.6,2.4}
{
    \draw[thick] (\x,0) -- (\x+0.8,0);
}

\node at (1.6,-0.6) {\scriptsize $i \;\;\; i+1 \;\;\; i+2 \;\;\; i+3 \;\;\; i+4$};

\node at (6,2.0) {\small (b) Hydrogen bond $i \to i+4$};

\foreach \x in {5,5.8,6.6,7.4,8.2}
{
    \fill[black] (\x,0) circle (0.08);
}

\foreach \x in {5,5.8,6.6,7.4}
{
    \draw[thick] (\x,0) -- (\x+0.8,0);
}

\draw[dashed, thick] (5,0.2) to[out=60,in=120] (8.2,0.2);

\node at (6.6,1.0) {\scriptsize H-bond};

\node at (12,2.0) {\small (c) Coarse-grained unit ($m=4$)};

\draw[thick, rounded corners] (10.5,-0.5) rectangle (13.5,0.5);

\node at (12,0) {\small segment};

\end{tikzpicture}

\caption{
Helix formation via hydrogen bonding. 
(a) Polymer chain at the monomer level.
(b) Formation of a hydrogen bond between residues $i$ and $i+4$, characteristic of $\alpha$-helical geometry.
(c) Coarse-grained description in which $m=4$ monomers are grouped into a single effective segment.
The case $m=4$ corresponds to $\alpha$-helical geometry, where hydrogen
bonds form between residues $i$ and $i+4$.
}
\end{figure}


\section{Theoretical Formulation}

The theory presented here is a minimal, coarse-grained description
based on an effective Hamiltonian. While the model has limitations, discussed
later, its simplicity allows a transparent description of spontaneous symmetry
breaking arising from collective interactions. This aspect is particularly
revealing and forms a central theme of the present work.

\subsection{Physical basis: periodic sticky interactions and coarse-graining}

We begin by recalling that helix formation in a linear polymer chain may arise from
specific attractive interactions between monomers separated by a fixed contour distance.
In particular, hydrogen bonding or sticker-like interactions often occur between sites
that are separated by a characteristic number of monomers ($m$), so that the interaction
distance along the contour is
\begin{equation}
\Delta s = m b,
\end{equation}
where $b$ is the monomer size.

For $\alpha$-helical polypeptides, intramolecular hydrogen bonding occurs
between residues separated by four positions along the backbone ($i \to i+4$),
as established by Pauling and Corey. In the present coarse-grained description,
this geometric constraint is incorporated by choosing $m=4$, although the
formalism applies to general $m$.

As discussed in our earlier work~\cite{Bagchi2026}, such periodic interactions can
stabilize helical conformations when the chain geometry allows monomers separated by
$\Delta s$ to come into spatial proximity. While the detailed geometric realization
depends on helix radius, pitch, and steric constraints, the essential requirement is
commensurability between interaction spacing and backbone geometry, enabling multiple
contacts to be satisfied simultaneously.

In the present work, we adopt a coarse-grained description in which each effective unit
represents a segment of $m$ monomers capable of forming a cooperative hydrogen-bonded
or sticky contact. The total number of such coarse-grained units is therefore
\begin{equation}
N_S = \frac{N_{\rm PC}}{m},
\end{equation}
where $N_{\rm PC}$ is the total number of physical monomers.

This coarse-graining is motivated by the fact that helix formation is intrinsically
cooperative: an isolated contact does not produce a stable helical structure, whereas
a sequence of correlated contacts can stabilize a locally ordered segment. The effective
degrees of freedom in the present theory therefore correspond not to individual monomers
but to these cooperative units.

\subsection{Discrete three-state description}

We consider a polymer chain of $N$ coarse-grained segments indexed by $i = 1, \ldots, N$.
Each segment is assigned a discrete variable
\begin{equation}
s_i \in \{0, +1, -1\},
\end{equation}
where
\begin{enumerate}[label=(\roman*)]
\item $s_i = 0$ \quad denotes the open (coil) state,
\item $s_i = +1$ \quad denotes a right-handed helical state,
\item $s_i = -1$ \quad denotes a left-handed helical state.
\end{enumerate}

It is convenient to introduce the auxiliary variable
\[
n_i = s_i^2 \in \{0,1\},
\]
which indicates whether a given segment is in the helical state, independent of
handedness. When $n_i = 1$, the chirality variable $\chi_i = s_i = \pm 1$
specifies the handedness.

It should be emphasized that the chirality variable $\chi_i = \pm 1$ is
defined at the level of a coarse-grained segment and does not represent
a strictly local property of a single monomer. Rather, it reflects the
sense of twist of a minimally cooperative helical unit spanning several
monomers. Consequently, well-defined chirality emerges only when
neighboring helical units are present and stabilized through cooperative
interactions.

Although right- and left-handed helices (RHH and LHH) are degenerate in bulk, their
juxtaposition along a single chain leads to a highly unfavorable
configuration due to geometric and torsional mismatch. Such RHH--LHH
junctions correspond to domain-wall defects with a large energetic cost.
In the regime of strong chiral coupling, these defects are suppressed,
and helical segments tend to exhibit uniform handedness.


\subsection{Free-energy functional}

We introduce an effective coarse-grained free-energy functional,
denoted by $H$, which plays the role of a Hamiltonian in the
statistical-mechanical description. Each segment can be either open
or helical with a definite chirality. The effective Hamiltonian is
written as
\begin{equation}
\begin{aligned}
H =&
\sum_{i=1}^{N} \left[\epsilon_i n_i - h_i (1 - n_i)\right] \\
&- K \sum_{i=1}^{N-1} n_i n_{i+1} \frac{1+\chi_i \chi_{i+1}}{2} \\
&- J \sum_{i=1}^{N-1} n_i n_{i+1} \chi_i \chi_{i+1}.
\end{aligned}
\label{eq:Hamiltonian}
\end{equation}

This Hamiltonian is motivated by lattice models of helix--coil transitions
and generalized spin models with internal states~\cite{ZimmBragg1959,LifsonRoig1961,Blume1966}.
It provides a minimal statistical-mechanical description of helix formation
that incorporates local energetics, cooperative stabilization, chirality,
and solvent-induced heterogeneity.

\vspace{0.3cm}

(i) The local energy term $\epsilon_i$ represents the effective free-energy cost of
forming a helical unit. It is a renormalized quantity that includes both
stabilizing contributions from hydrogen bonding (or sticky interactions)
and opposing effects such as conformational entropy loss, torsional
constraints, and local bending penalties. The local free-energy parameter
may be written schematically as
\[
\epsilon_i = \epsilon_{\mathrm{HB}} + \epsilon_{\mathrm{torsion}} + \epsilon_{\mathrm{bend}},
\]
where $\epsilon_{\mathrm{HB}}$ denotes hydrogen-bond stabilization, while
$\epsilon_{\mathrm{torsion}}$ and $\epsilon_{\mathrm{bend}}$ represent
conformational penalties. In general, an isolated helical unit is unstable,
so that $\epsilon_i + h_i > 0$.

(ii) The field $h_i$ accounts for solvent-induced stabilization of the open
state. In aqueous systems, this reflects hydrogen bonding of the polymer
with water, which locally favors coil configurations. Spatial variations
in $h_i$ introduce heterogeneity and act as an effective random field.

(iii) The chirality variable $\chi_i = \pm 1$ denotes the handedness of a
helical unit at site $i$. For open (coil) sites ($n_i = 0$), $\chi_i$ is
undefined but does not contribute to the energy due to the multiplicative
factor $n_i$.

(iv) The cooperative interaction term proportional to $K$ captures the
essential physics that helix formation is not purely local but collective.
Importantly, in the present formulation, \emph{cooperative stabilization is
conditional on chiral alignment}. The factor $(1+\chi_i\chi_{i+1})/2$
ensures that neighboring helical units contribute to stabilization only
when they possess the same handedness. Thus, aligned segments reinforce
each other, whereas mismatched chiralities do not contribute to cooperative
stabilization. This reflects the geometric requirement that periodic
contacts can be simultaneously satisfied only when local twist is coherent.

(v) The chiral coupling $J > 0$ further penalizes mismatches in handedness
between neighboring helical units. While the $K$ term controls the
existence of cooperative stabilization, the $J$ term controls the energetic
cost of domain walls. Together, these terms ensure that extended helical
segments tend to adopt uniform chirality, while suppressing RHH--LHH
interfaces.

\vspace{0.3cm}

The parameters $(\epsilon_i, K, J)$ are effective free-energy quantities
that emerge from coarse-graining over microscopic degrees of freedom.
They should not be interpreted as bare interaction energies, but as
renormalized parameters that encode local structure, solvent effects,
and collective interactions.

Equation~(\ref{eq:Hamiltonian}) makes explicit that helix formation is
intrinsically a cooperative phenomenon: an isolated helical unit is
unstable, but a cluster of neighboring units can become stabilized once
cooperative interactions overcome the local free-energy cost. The crucial
feature of the present formulation is that this cooperativity is not purely
geometric, but also requires chiral compatibility, leading naturally to
symmetry breaking and the emergence of uniformly handed helical domains.

The present formulation is intended as a coarse-grained, minimal
description of helix formation. The variables $(n_i,\chi_i)$ do not
represent microscopic degrees of freedom, but rather effective local
states that encode the presence of a helical segment and its handedness.
The parameters $(\epsilon_i, K, J, h_i)$ therefore represent effective
free-energy quantities incorporating hydrogen bonding, torsional
constraints, and solvent interactions. The goal of this model is not
to reproduce microscopic detail, but to capture the essential physics
of cooperativity, symmetry breaking, and dynamical fluctuations in
helix formation.

In the present work, the explicit chirality variables mainly serve to establish the cooperative and symmetry-breaking framework. However, most numerical estimates developed later are performed in an effective occupancy representation where the dominant effect of chirality is absorbed into renormalized cooperative parameters.

\section{Thermodynamic considerations from statistical mechanics}

The coarse-grained Hamiltonian introduced in Eq.~(\ref{eq:Hamiltonian})
defines a one-dimensional lattice model in which each site can exist in
one of three effective states: an open (coil) state ($n_i = 0$) or a
helical state ($n_i = 1$) with chirality $\chi_i = \pm 1$. Thus, the
local state space may be viewed as $s_i \in \{0, +, -\}$, corresponding
to a three-state model with constrained interactions.

From a statistical-mechanical perspective, the thermodynamics of this
system is governed by the partition function
\begin{equation}
Z = \sum_{\{n_i,\chi_i\}} e^{-\beta H},
\end{equation}
where the sum runs over all configurations of occupancy and chirality.
Because the Hamiltonian contains only nearest-neighbor interactions,
the model can, in principle, be analyzed using a transfer-matrix
formalism, as is standard for one-dimensional lattice systems with a
finite number of local states.

The thermodynamic behavior is determined by the competition between
local energetic penalties and cooperative interactions. For an isolated
site, the free-energy cost of forming a helical unit is $\epsilon_i + h_i$,
which is typically positive due to conformational entropy loss and
solvent stabilization of the coil state. However, in the presence of
neighboring helical segments, the cooperative terms proportional to $K$
and $J$ reduce this cost, favoring the formation of extended helical
regions.

At zero temperature, the system adopts the configuration that minimizes
the total energy. Depending on the relative magnitudes of the parameters,
two limiting states may be identified: a coil-dominated state, favored
when $\epsilon_i + h_i$ is large, and a uniformly helical state with a
definite chirality, favored when the cooperative interactions are
sufficiently strong. Thus, the model exhibits a competition between
disordered and ordered configurations.

At finite temperature, however, the situation is qualitatively different.
As a one-dimensional system with short-range interactions, the model does
not exhibit a true thermodynamic phase transition, but rather a
cooperative crossover between coil-rich and helix-rich regimes. Thermal
fluctuations lead to the formation of finite helical segments separated
by coil regions, consistent with the picture of dynamically fluctuating
``living helices'' developed in this work.

Although an exact analytical treatment is formally possible via a
transfer-matrix approach, the coupling between occupancy and chirality
leads to algebraically involved expressions that obscure the underlying
physics. For this reason, the present work focuses on a physically
transparent free-energy description and its associated kinetics, which
capture the essential features of helix formation, stability, and fluctuation.

While the thermodynamic description characterizes the equilibrium
properties of the system, the formation and evolution of helical
segments is intrinsically a dynamical process. We therefore next
introduce a stochastic description consistent with the Hamiltonian.


\section{Stochastic dynamics with the coarse-grained Hamiltonian}

To connect the coarse-grained Hamiltonian to the kinetics of helix
formation, growth, and decay, we introduce a stochastic description at
the level of individual sites. The state of the system at time $t$ is
specified by the joint probability distribution
\begin{equation}
P(\{n_i,\chi_i\},t),
\end{equation}
where $n_i \in \{0,1\}$ denotes the local occupancy (coil or helix) and
$\chi_i=\pm 1$ denotes the chirality of a helical unit. The chirality
variable is defined only when $n_i=1$.

The stochastic description employed here is based on the standard
master-equation framework for Markov processes, widely used to describe
activated dynamics, random walks, and nucleation phenomena in statistical
physics. In particular, the evolution in configuration or size space can
be viewed as a birth–death process, and quantities such as lifetimes and
formation times are naturally formulated as mean first-passage-time (MFPT)
problems. These approaches have a long history in the study of chemical
kinetics, diffusion processes, and stochastic dynamics
\cite{BagchiNESM,VanKampen,Gardiner,Risken,Redner}.

\subsection{Master equation}

We consider a local, single-site dynamics in which transitions occur
only between coil and helical states, i.e., $n_i=0 \leftrightarrow 1$.
The time evolution of the probability distribution is governed by the
master equation
\begin{equation}
\frac{dP}{dt} = \sum_i \sum_{n_i' \neq n_i}
\Big[
W_i(n_i' \to n_i \mid \chi)\, P(\ldots,n_i',\chi,\ldots,t)
-
W_i(n_i \to n_i' \mid \chi)\, P(\ldots,n_i,\chi,\ldots,t)
\Big],
\label{eq:master_full}
\end{equation}
where $W_i$ are the transition rates and the notation $(\ldots)$
indicates that all other variables are unchanged.
\noindent

\subsection{Role of chirality}

In the present formulation, the chirality variable $\chi_i$ is not
treated as an independently evolving dynamical degree of freedom.
Instead, it is generated during helix formation and remains fixed
thereafter. Thus, no direct transitions of the form
$\chi_i \to -\chi_i$ are included. The stochastic dynamics acts only
on the occupancy variable $n_i$, while the influence of chirality
enters implicitly through the local energy changes.

When a site transitions from the coil to the helical state
($n_i=0 \to 1$), the chirality $\chi_i=\pm1$ is assigned according
to the local energetic bias imposed by neighboring sites, ensuring
consistency with detailed balance. When a helical site melts
($n_i=1 \to 0$), the chirality becomes irrelevant.
 An explicit expression for $\Delta E_i$ is given
below in Eq.~(\ref{eq:deltaE}).

The choice of transition rates is not unique, but is constrained by the
requirement of detailed balance with respect to the underlying Hamiltonian.
Specifically, the ratio of forward and backward transition rates must satisfy
\begin{equation}
\frac{W_i(s_i \to s_i')}{W_i(s_i' \to s_i)}
= \exp\big[-\beta \Delta E_i(s_i \to s_i')\big],
\end{equation}
where $\Delta E_i$ is the local free-energy change associated with the
transition. This condition ensures that the stochastic dynamics relaxes
to the correct equilibrium distribution.

A convenient and widely used choice satisfying this condition is the Glauber form
\begin{equation}
W_i(s_i \to s_i') =
\frac{1}{\tau_0}
\frac{1}{1 + \exp\big[\beta \Delta E_i(s_i \to s_i')\big]},
\end{equation}
originally introduced in the context of kinetic Ising models \cite{Glauber1963}.
It is straightforward to show that this form satisfies the principle of detailed balance.
This form has the advantage of being symmetric, bounded, and physically
transparent, with energetically favorable transitions occurring with
higher probability while maintaining detailed balance.

We emphasize that other choices of rates (e.g., Metropolis dynamics) would
lead to the same equilibrium behavior, although dynamical properties such
as timescales may differ. The present choice provides a minimal and
analytically convenient representation of the stochastic dynamics.

This stochastic dynamics describes the local nucleation and melting
of helical segments through transitions in the occupancy variable
$n_i$, while the chirality of newly formed segments is determined by
the surrounding configuration. Because chirality is not allowed to
flip dynamically, left- and right-handed helical domains emerge
through nucleation events and persist unless destroyed by melting.

The formulation thus provides a physically consistent link between
the coarse-grained Hamiltonian and stochastic evolution. In the
following sections, we focus on a further coarse-grained description
in terms of the helix size $n$ and analyze the associated kinetics
using mean first-passage-time methods.

\section{Local bistability and cooperative nucleation}
We recall that each site $i$ represents a coarse-grained unit corresponding
to a segment of $m$ monomers capable of forming a cooperative sticky or
hydrogen-bonded contact. The local variables therefore describe the state
of such effective units rather than individual monomers.

The effective local free energy at site $i$ depends on its neighbors.
Here, $\epsilon_i$ represents the effective local free-energy cost of forming
a helical unit of size $m$, incorporating the balance between sticky (or
hydrogen-bonding) stabilization and opposing contributions such as
conformational entropy loss, bending, and torsional constraints.

For a given configuration of neighboring sites, the free-energy change associated with
converting a site from open to helical can be written as

\begin{equation}
\begin{split}
\Delta E_i = \epsilon_i + h_i 
- K \Big[
n_{i-1}\frac{1+\chi_{i-1}\chi_i}{2}
+ n_{i+1}\frac{1+\chi_{i+1}\chi_i}{2}
\Big] \\
- J (n_{i-1}\chi_{i-1} + n_{i+1}\chi_{i+1}) \chi_i .
\end{split}
\label{eq:deltaE}
\end{equation}

Each term in Eq.~(\ref{eq:deltaE}) has a clear physical interpretation.The first two terms, $\epsilon_i + h_i$, represent the intrinsic free-energy cost of
forming a helical unit at site $i$. The parameter $\epsilon_i$ includes the balance
between hydrogen-bond stabilization and conformational penalties such as torsion
and bending, while $h_i$ accounts for solvent-induced stabilization of the open
(coil) state. When $\epsilon_i + h_i > 0$, an isolated helical unit is energetically
unfavorable.

In the present formulation, cooperative stabilization is conditional on
chiral compatibility, and is included in the third term. A neighboring helical unit contributes to the
cooperative term only when it has the same handedness as the newly
formed unit. This is implemented through the factor
$(1+\chi_{i\pm1}\chi_i)/2$, which equals unity for matching chirality
and vanishes for mismatched chirality.
 Each adjacent helical site lowers the energy of
forming a helix at site $i$ by an amount $K$. Thus, if both neighbors are helical,
the energy is reduced by $2K$, reflecting the fact that local structural ordering
and the satisfaction of multiple periodic interactions reinforce one another.

The fourth term,
\[
- J (n_{i-1}\chi_{i-1} + n_{i+1}\chi_{i+1}) \chi_i,
\]
encodes chiral alignment independently of the cooperative stabilization.
This contribution is active only when neighboring sites are in the
helical state ($n_{i\pm1}=1$), since the chirality variables are defined
only for helical units. In that case, the interaction reduces to an
effective Ising-like coupling between neighboring chirality variables.

If the chirality at site $i$ matches that of a neighboring helical site,
the energy is lowered by an amount $J$, whereas a mismatch incurs an
energetic penalty of the same magnitude. Thus, this term does not
contribute to the existence of cooperativity itself, but instead
controls the energetic cost of chiral mismatches within an already
formed helical segment.

Consequently, the $J$ term suppresses domain walls between left- and
right-handed segments and promotes the formation of extended regions
with uniform handedness.

Taken together, Eq.~(\ref{eq:deltaE}) shows that helix formation is intrinsically
a collective phenomenon: while an isolated helical unit is unfavorable, the presence
of neighboring helical segments both stabilizes its formation (through $K$) and
enforces coherent handedness (through $J$).

Thus, cooperativity and chirality are coupled but not identical: $K$
controls the existence of cooperative stabilization, while $J$
controls the energetic selection of a specific handedness.

The above expression for $\Delta E_i$ makes it clear that helix formation
cannot proceed via the stabilization of a single site. Instead, the
energetic gain from cooperative interactions must overcome the local
free-energy cost $\epsilon_i + h_i$. This requires the formation of a
cluster of neighboring helical sites.

A simple estimate of the stability of a helical cluster may be obtained
by considering a contiguous segment of length $n$ embedded in an open
background. Neglecting disorder for the moment, the total free-energy
change associated with forming such a cluster can be written schematically as
\begin{equation}
\Delta E(n) \simeq n(\epsilon + h) - (K+J) (n-1),
\end{equation}
where the first term represents the local cost of forming $n$ helical
units and the second term represents the cooperative stabilization due
to $(n-1)$ favorable nearest-neighbor contacts.

It is convenient to define the effective bulk free-energy gain per
helical unit as
\begin{equation}
\Delta f = (K+J) - (\epsilon + h) .
\end{equation}
In terms of this quantity, helix growth becomes favorable when
$\Delta f > 0$.

For a uniformly handed helical segment, the conditional cooperative term and chiral alignment term act together, leading to an effective stabilization $K_{eff} = K + J$.

That is, helical growth becomes favorable when
\begin{equation}
\epsilon + h - K - J < 0 .
\end{equation}
while small clusters remain unstable due to the effective boundary
free-energy cost associated with the loss of cooperative contacts
at the edges of the segment.

The existence of a boundary cost implies a minimal cluster size
$n_{\mathrm{min}}$ below which helical segments are unstable and decay
rapidly. Thus, nucleation requires the formation of a finite segment
whose size is sufficient for cooperative stabilization to overcome the
local energetic cost.

The chiral alignment term proportional to $J$ introduces an additional
constraint. Since this term couples neighboring chirality variables, it
becomes effective only when at least two adjacent helical sites are
present. Consequently, a well-defined handedness cannot emerge at the
level of an isolated bond and instead requires the buildup of a minimally
cooperative segment. Chirality is therefore not a purely local property
but an emergent feature of correlated multi-site configurations.

These considerations show that helix nucleation is intrinsically
multistep: initial bond formation produces a locally constrained
configuration, but additional correlated bonds are required both to
stabilize the helical state and to establish a coherent chirality. This
multistep nature of nucleation will be analyzed in detail in a later
section.

\begin{figure}[t]
\centering
\begin{tikzpicture}[scale=1.2]

\node at (0,2.2) {\small Right-handed ($\chi=+1$)};

\draw[thick] (0,0) .. controls (0.5,0.5) and (-0.5,1.0) .. (0,1.5)
             .. controls (0.5,2.0) and (-0.5,2.5) .. (0,3);

\draw[dashed] (0,-0.2) -- (0,3.2);

\node at (4,2.2) {\small Defect};

\draw[thick] (4,1.6) .. controls (4.5,2.0) and (3.5,2.4) .. (4,2.8);

\draw[thick] (4,1.6) -- (4.3,1.3) -- (3.7,1.0);

\draw[thick] (4,1.0) .. controls (4.5,0.6) and (3.5,0.2) .. (4,-0.2);

\draw[dashed] (4,-0.5) -- (4,3.2);

\node at (4,-1.0) {\scriptsize $\sim (K+J)$};

\node at (8,2.2) {\small Left-handed ($\chi=-1$)};

\draw[thick] (8,0) .. controls (-0.5+8,0.5) and (0.5+8,1.0) .. (8,1.5)
             .. controls (-0.5+8,2.0) and (0.5+8,2.5) .. (8,3);

\draw[dashed] (8,-0.2) -- (8,3.2);

\end{tikzpicture}

\caption{
Right- and left-handed helical segments ($\chi=\pm1$) are degenerate in energy.
A junction between opposite chiralities creates a domain wall (defect) with an energetic penalty
controlled by the loss of cooperative stabilization and chiral alignment, with
an energy cost $\sim (K+J)$, favoring extended regions of uniform handedness.
}
\end{figure}
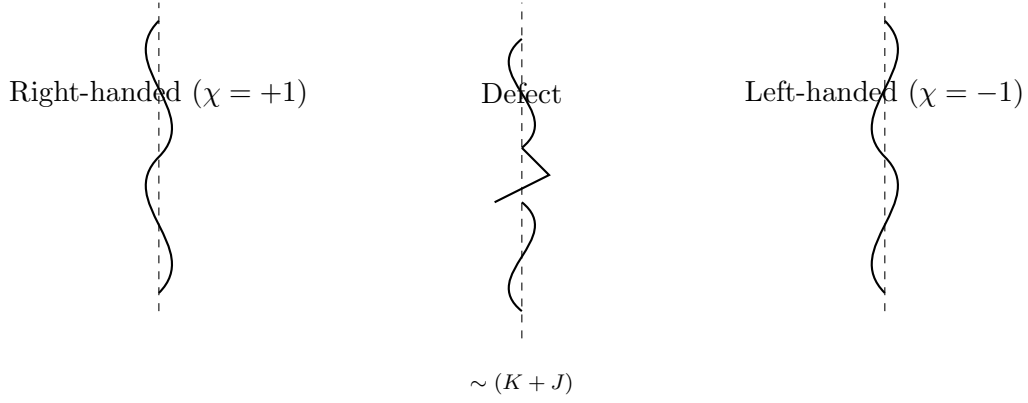

\subsection{Stochastic dynamics}

The time evolution of the system is described by stochastic transitions
between the three local states. We consider single-site dynamics of the form
\begin{equation}
s_i : 0 \rightleftharpoons \pm 1,
\end{equation}
with transition rates satisfying detailed balance.

The rate $W$ for transition $s_i \to s_i'$ is given in Section 4.


Direct chirality flips ($+1 \leftrightarrow -1$) may occur either
through higher-order processes or via the intermediate open state,
\begin{equation}
+1 \to 0 \to -1.
\end{equation}

The probability distribution $P(\{s_i\},t)$ evolves according to the
master equation
\begin{equation}
\frac{dP}{dt} =
\sum_i \sum_{s_i'} \left[
W(s_i' \to s_i) P(\dots,s_i',\dots,t)
- W(s_i \to s_i') P(\dots,s_i,\dots,t)
\right].
\end{equation}

\subsection{Helical domains and boundary dynamics}

A helical segment of length $n$ is defined as a contiguous region
\begin{equation}
n_i = 1 \quad \text{for} \quad i = \ell, \dots, \ell+n-1,
\end{equation}
bounded by open sites ($n_{\ell-1}=0$, $n_{\ell+n}=0$).

The free energy of such a segment can be written approximately as
\begin{equation}
F(n) = 2\sigma - n \Delta f + \delta F_{\text{dis}},
\label{eq:Fn}
\end{equation}
where:

\begin{enumerate}
\item $\sigma$ is the effective boundary (domain-wall) free energy,
arising from broken cooperative interactions,
\item $\Delta f$ is the bulk free-energy difference between helix and coil,
\item $\delta F_{\text{dis}}$ represents corrections due to spatial
heterogeneity in $\epsilon_i$ and $h_i$.
\end{enumerate}

The form (\ref{eq:Fn}) provides a coarse-grained representation of the
microscopic Hamiltonian. 
In particular, the cooperative interactions encoded in $K$ and $J$ generate an effective bulk stabilization $\Delta f$, while the loss of cooperative contacts at the boundaries gives rise to the interfacial free-energy cost $\sigma$. This interfacial cost reflects the fact that boundary units do not benefit from the full cooperative stabilization present in the interior of a helical segment.

The dynamics of a helical segment is governed by stochastic motion of its
boundaries. Growth and shrinkage occur through local transitions at the
edges,
\begin{equation}
n \rightarrow n \pm 1,
\end{equation}
with rates determined by the corresponding changes in $F(n)$.

\subsection{Theoretical objectives}

The theoretical framework developed above allows us to address several
key questions:

(i) What determines the lifetime $\tau(n)$ of a helical segment of length $n$?

(ii) Under what conditions does the system exhibit marginally stable,
finite helices?

(iii) How do cooperative interactions and solvent-induced fluctuations
modify the effective free-energy landscape?

(iv) How does the stochastic motion of domain boundaries lead to migration
of helical segments along the chain?

In the following sections, we develop analytical expressions for these
quantities by treating the growth and shrinkage of helical domains as a
stochastic birth--death process in the free-energy landscape defined by
Eq.~(\ref{eq:Fn}), and by analyzing the resulting coupled dynamics of
segment size and position.


\section{Lifetime of a Helical Segment: Birth--Death Dynamics in Size Space}

We now formulate the lifetime of a helical segment as a first-passage
problem in the size variable $n$. Here $n$ denotes the number of contiguous
coarse-grained helical units in a segment. The segment grows or shrinks
through local events at its boundaries. Thus, the elementary processes are
\[
n \to n+1, \qquad n \to n-1 .
\]
We denote by $u_n$ the rate for growth from size $n$ to size $n+1$, and
by $w_n$ the rate for shrinkage from size $n$ to size $n-1$. These rates
will in general depend on $n$, because the free energy of the segment may
depend nonlinearly on its size.

Let $P(n,t)$ be the probability that the helical segment has size $n$ at
time $t$. The size dynamics is then described by the birth--death master
equation
\begin{equation}
\frac{dP(n,t)}{dt}
=
u_{n-1}P(n-1,t)
+
w_{n+1}P(n+1,t)
-
\left(u_n+w_n\right)P(n,t).
\label{eq:master_size_lifetime}
\end{equation}
The rates are constrained by detailed balance with respect to the effective
free energy $F(n)$:
\begin{equation}
\frac{u_n}{w_{n+1}}
=
\exp\left[-\beta\left(F(n+1)-F(n)\right)\right].
\label{eq:db_un_wn}
\end{equation}
This relation ensures that the stochastic dynamics in size space is
consistent with the coarse-grained thermodynamics.

For the linear free-energy model, we take
\begin{equation}
F(n)=2\sigma-n\Delta f ,
\label{eq:F_linear_lifetime}
\end{equation}
where $2\sigma$ is the cost of creating two helix--coil boundaries and
$\Delta f$ is the net free-energy gain per helical unit. In this case
\[
F(n+1)-F(n)=-\Delta f ,
\]
and Eq.~(\ref{eq:db_un_wn}) gives
\begin{equation}
\frac{u_n}{w_{n+1}}=\exp(\beta \Delta f).
\label{eq:db_linear}
\end{equation}
Since the free-energy difference is independent of $n$, the rates become
independent of size. We therefore write
\[
u_n=u, \qquad w_n=w ,
\]
so that
\begin{equation}
\frac{u}{w}=\exp(\beta \Delta f).
\label{eq:u_w_ratio}
\end{equation}
Here $u$ is the constant growth rate and $w$ is the constant shrinkage
rate in the linear free-energy approximation.

The lifetime $\tau(n)$ is defined as the mean first-passage time for a
helical segment of initial size $n$ to disappear, i.e., to reach the
absorbing state $n=0$. 
This backward equation is standard in the theory of stochastic
processes and first-passage problems~\cite{VanKampen,Gardiner,Redner}.
Here the backward equation for the mean lifetime is
\begin{equation}
-1
=
u_n\left[\tau(n+1)-\tau(n)\right]
+
w_n\left[\tau(n-1)-\tau(n)\right],
\qquad n\ge 1,
\label{eq:backward_lifetime_general}
\end{equation}
with absorbing boundary condition
\begin{equation}
\tau(0)=0.
\label{eq:absorbing_lifetime}
\end{equation}

For the linear free-energy model, where $u_n=u$ and $w_n=w$, this becomes
\begin{equation}
-1
=
u\left[\tau(n+1)-\tau(n)\right]
+
w\left[\tau(n-1)-\tau(n)\right].
\label{eq:backward_lifetime_linear}
\end{equation}
In the unstable regime, $w>u$, shrinkage is favored and the solution
that remains finite for large $n$ is
\begin{equation}
\tau(n)=\frac{n}{w-u}.
\label{eq:tau_linear_result}
\end{equation}
Using Eq.~(\ref{eq:u_w_ratio}), this can also be written as
\begin{equation}
\tau(n)=
\frac{n}{w\left(1-e^{\beta \Delta f}\right)},
\qquad \Delta f<0.
\label{eq:tau_linear_db}
\end{equation}

Equation~(\ref{eq:tau_linear_db}) provides the baseline result for the
lifetime of an unstable helical segment. The lifetime grows linearly with
the initial size $n$, because the segment disappears through successive
boundary melting events. It also increases strongly as $\Delta f$ approaches
zero from below. In this limit, growth and shrinkage nearly balance, and
the helix becomes marginally stable. Thus, even the linear model already
reveals the onset of a living-helix regime near $\Delta f=0$.

However, the linear free energy cannot describe stable finite helices. If
$\Delta f<0$, helices shrink and disappear; if $\Delta f>0$, the linear
model predicts indefinite growth. Therefore, finite long-lived helices
require additional nonlinear contributions to $F(n)$. These terms are
introduced in the next section, where $u_n$ and $w_n$ become explicitly
size-dependent through Eq.~(\ref{eq:db_un_wn}), leading to activated
lifetimes and a preferred finite helix size.


\section{Finite Helices and Activated Lifetimes: Nonlinear Free-Energy Landscape}

The linear free-energy model discussed in the previous section provides a
useful baseline description of helix dynamics, but it does not account for
the existence of stable or metastable finite helical segments. In that
model, helices either shrink monotonically when $\Delta f < 0$, or grow
without bound when $\Delta f > 0$. Neither scenario captures the experimentally
observed behavior of finite, fluctuating helical segments.

To describe such behavior, it is necessary to incorporate nonlinear
contributions to the free energy that penalize excessive growth and
stabilize helices of finite size. These nonlinearities arise naturally
from stiffness, torsional strain, packing constraints, and other geometric
effects that become increasingly important as the size of the helical
segment grows.

\subsection{Nonlinear free-energy model}

We therefore consider a generalized free-energy function of the form
\begin{equation}
F(n) = 2\sigma - n\Delta f + U(n),
\label{eq:F_nonlinear}
\end{equation}
where $U(n)$ represents a nonlinear contribution that increases with $n$.
Physically, $U(n)$ encodes penalties associated with bending rigidity,
torsional frustration, excluded-volume effects, and deviations from optimal
helical geometry.

The nonlinear term characterized by $\alpha$ arises naturally from
elastic and geometric constraints in semiflexible chains. In a
coarse-grained description, the free energy may be expressed schematically
as $F(n) = 2\sigma - n\Delta f + \alpha n^2 + \cdots$, where the quadratic
term represents an effective penalty that increases with helix size.
Physically, this term incorporates contributions from bending rigidity,
torsional strain, and packing constraints, all of which oppose unlimited
growth of the helical segment. Similar nonlinear size-dependent penalties
are well known in the theory of semiflexible polymers and in the energetics
of DNA condensation and toroidal structures, where elastic costs compete
with cohesive interactions to produce finite, stable morphologies
\cite{DoiEdwards,KratkyPorod,Bloomfield1997,UbbinkOdijk1999}. In this
sense, the parameter $\alpha$ plays the role of an effective stiffness,
governing the crossover from cooperative growth to geometric frustration
and thereby stabilizing helices of finite size.

For small $n$, the cooperative term $-n\Delta f$ dominates, favoring growth
of the helical segment. However, for sufficiently large $n$, the nonlinear
term $U(n)$ becomes significant and counteracts further growth. This
competition leads to a non-monotonic free-energy profile.

\subsection{Emergence of a finite preferred size}

The condition for a stationary point of $F(n)$ is
\begin{equation}
\frac{dF}{dn} = -\Delta f + \frac{dU}{dn} = 0.
\label{eq:stationary_condition}
\end{equation}
This defines a characteristic size $n_c$ at which the free energy is
extremal. Depending on the form of $U(n)$, this extremum may correspond
to a local minimum, indicating a metastable helical segment of finite
size.

The existence of such a minimum implies that helices are neither unstable
nor indefinitely growing, but instead exist as finite, dynamically
fluctuating structures. This provides a natural explanation for the
“living helix” regime observed in experiments and simulations.


\subsection{Activated decay from the finite helical state}

For the quadratic free-energy model used below, the finite helical state corresponds
to a minimum. The activated character of the lifetime should
therefore be understood as an effective first-passage process associated with escape
from the finite helical segment toward the open state. In a discrete description, this
escape involves successive boundary-melting events and the loss of cooperative contacts.
The relevant free-energy scale is the depth of the finite-size minimum relative to the
boundary or open-chain reference.

This barrier controls the lifetime of the helical segment. In contrast
to the linear model, where decay occurs through a simple biased random
walk, the presence of a barrier leads to activated dynamics. The helix
must undergo a rare fluctuation to cross the barrier before it can decay
to the coil state.

\subsection{Connection to stochastic dynamics}

The stochastic description introduced in Section 6 remains fully valid
in the nonlinear regime. However, the transition rates $u_n$ and $w_n$
now acquire an explicit dependence on $n$ through the detailed-balance
condition
\begin{equation}
\frac{u_n}{w_{n+1}} =
\exp\big[-\beta (F(n+1) - F(n))\big].
\label{eq:db_nonlinear}
\end{equation}

Thus, the nonlinear free-energy landscape directly determines the local
growth and shrinkage rates. Near the minimum at $n_c$, the rates become
nearly balanced, leading to slow fluctuations in size. Near the barrier,
the rates become strongly biased, governing the escape process.

\subsection{Activated lifetime}

The lifetime of a helical segment in this regime can be estimated using
the mean first-passage framework developed in Section 6. In the presence
of a barrier, the lifetime acquires an activated form,
\begin{equation}
\tau \sim \tau_0 \exp(\beta \Delta F^\ddagger),
\label{eq:activated_lifetime}
\end{equation}
where $\tau_0$ is a characteristic microscopic timescale determined by
the local transition rates.

This exponential dependence reflects the fact that decay requires crossing
a free-energy barrier, and therefore occurs through rare fluctuations.
The lifetime can thus become very large even for moderate barrier heights.

\subsection{Physical interpretation}

The nonlinear free-energy landscape provides a microscopic picture
of helix stability. The competition between local energetic penalties
and cooperative stabilization leads to the emergence of a preferred size,
while stiffness and geometric constraints prevent unlimited growth.

As a result, helices behave as finite, marginally stable objects that
continuously fluctuate in size and can be created or destroyed through
activated processes. This behavior underlies the concept of “living
helices,” in which structure and dynamics are intimately coupled.


\section{Migration of helical segments}

A helical segment is not stationary but undergoes stochastic motion along the chain.
It is important to note that a finite helical segment possesses two
boundaries, and its dynamics involves stochastic motion of both edges.
In addition to growth and shrinkage, correlated motion of the two
boundaries can lead to translation of the entire helical segment along
the chain. This mechanism is closely analogous to the dynamics of
localized excitations in polaron models and to bubble migration in DNA,
where the motion of domain boundaries gives rise to diffusive transport
of the excitation. In the present work, we focus primarily on size
fluctuations, while a more complete treatment of coupled size and
position dynamics will be developed in future work.

Let the segment occupy $(\ell,r)$, with center
\begin{equation}
x = \frac{\ell+r}{2}, \quad n=r-\ell+1.
\end{equation}

Each elementary boundary event produces a coupled change in both the
length and position of the segment. In particular, growth or melting at
either boundary shifts the center of the segment by $\pm 1/2$ lattice
units. Since these events occur stochastically with rates $u_n$ and $w_n$,
the center undergoes a random walk generated by a sequence of such
increments.

The mean-square displacement of the center after a short time interval
$\Delta t$ is therefore
\begin{equation}
\langle (\Delta x)^2 \rangle
= \frac{1}{4} (u_n + w_n)\, \Delta t,
\end{equation}
where the factor $1/4$ arises from the squared displacement
$(\Delta x = \pm 1/2)^2$ associated with each elementary event.

Taking the continuum limit, we obtain the diffusion coefficient
\begin{equation}
D_x(n) = \frac{u_n + w_n}{4}.
\end{equation}

Boundary fluctuations generate coupled dynamics in $(x,n)$.
Each elementary event shifts the center by $\pm 1/2$, leading to
diffusive motion.

\subsection{Migration length}

The distance traveled during the lifetime is
\begin{equation}
\ell_{\mathrm{mig}}(n) \sim \sqrt{2 D_x(n)\,\tau(n)}.
\end{equation}


We arrive at the following picture: (i)Helices are finite due to nonlinear penalties,
(ii)they are long-lived due to an activation barrier, and
(iii)they migrate along the chain via boundary fluctuations.

This behavior is qualitatively distinct from classical nucleation,
where clusters evolve only in order-parameter space.

\section{Relation to DNA bubble dynamics}

The present description bears strong analogy to bubble formation and
migration in DNA duplexes. In particular, water-mediated stabilization
of open states and the role of torsional constraints have been emphasized
in recent studies \cite{Mondal2024,Dasanna2013,Sicard2014}. We also refer to the classic articles in the area of DNA helix dynamics
\cite{AltanBonnet2003,Peyrard2004,PolandScheraga1966}

These works show that localized defects (bubbles) can migrate along the
DNA contour and exhibit finite lifetimes, reinforcing the physical picture
developed here for finite helices.

\subsection{Effect of solvent-induced random fields}

The interaction of the polymer with the surrounding solvent, particularly
water, introduces spatial heterogeneity in the stability of the helical
state. Local hydrogen bonding with solvent molecules can stabilize the
open configuration, effectively acting as a site-dependent field that
biases the helix--coil balance.

To incorporate this effect, we write the free energy as
\begin{equation}
F(n)=2\sigma - \sum_{i=1}^n (\Delta f - \eta_i) + \alpha n^2,
\end{equation}
where $\eta_i$ represents a local random field arising from solvent
fluctuations.

Upon coarse-graining, this leads to
\begin{equation}
F(n)=2\sigma - n\Delta f_{\mathrm{eff}} + \alpha n^2 + \delta(n),
\end{equation}
where $\Delta f_{\mathrm{eff}}=\Delta f - \bar{\eta}$ is the renormalized
driving force and $\delta(n)$ represents fluctuations.

The presence of the random field reduces the effective helix stability,
shifts the preferred size $n_c$, and introduces spatial variability in
both the lifetime and migration of helical segments.


\section{Kinetic Origin of Chirality and Multistep Nucleation}

The emergence of helicity in polymer chains is intrinsically a dynamical and
multistep process. In the absence of any explicit chiral bias in the underlying
Hamiltonian, the system is symmetric under inversion of handedness, and the
selection of a specific helical sense must arise through spontaneous symmetry
breaking during the nucleation process.

A key point is that helix nucleation cannot be viewed as a single-step transition
from the open state to a fully formed helical segment. Instead, it proceeds through
a sequence of intermediate states with progressively increasing structural
constraint and cooperativity.

\subsection{Pre-nucleation: formation of the first bond}

The formation of the first hydrogen bond between distant segments of the chain
constitutes the initial step, $0 \rightarrow 1$.
This step is dominated by a substantial loss of conformational entropy due to
the restriction of torsional and bending degrees of freedom. The resulting state
is a locally constrained configuration, but it does not yet possess any well-defined
chirality. In this sense, it represents a pre-nucleus.

\subsection{Emergence of chiral bias: second bond formation}

The formation of a second hydrogen bond introduces correlations between adjacent
segments, $1 \rightarrow 2^{*}$,
where $2^{*}$ denotes a partially organized state. At this stage, geometric
constraints begin to induce a local twist, leading to an incipient chiral bias.
However, the configuration remains nearly degenerate with respect to left- and
right-handed arrangements, and thermal fluctuations can still interconvert these
states. Thus, the second bond does not uniquely determine the handedness but
instead creates a dynamically fluctuating, chirality-biased intermediate.

\subsection{Selection of handedness: formation of a minimal helical nucleus}

A robust distinction between left- and right-handed helices emerges only upon
the formation of a minimally cooperative motif involving at least three correlated
bonds, $2^{*} \rightarrow \pm$,
where $+$ and $-$ denote right- and left-handed helices, respectively. At this
stage, the combined effects of torsional constraints and cooperative interactions
stabilize a specific sense of twist, and the system undergoes spontaneous symmetry
breaking.

This minimal nucleus may be viewed as the smallest segment for which the chiral
alignment term in the Hamiltonian,
\begin{equation}
- J \sum_i n_i n_{i+1} \chi_i \chi_{i+1},
\end{equation}
becomes effective in enforcing a uniform handedness. Once this nucleus is formed,
subsequent growth proceeds cooperatively, propagating the selected chirality along
the chain.

\subsection{Effective nucleation kinetics}

The overall nucleation process is therefore inherently multistep and may be
schematically represented as $0 \rightarrow 1 \rightarrow 2^{*} \rightarrow \pm$.
The effective nucleation rate is controlled by the sequential occurrence of these
steps and may be expressed, at a schematic level, as
$k_{\mathrm{nuc}} \sim k_1 \, k_2 \, k_3$,
where $k_1$ is the rate of initial bond formation, $k_2$ characterizes the formation
of the correlated intermediate, and $k_3$ corresponds to the symmetry-breaking
selection of chirality.

This multistep mechanism highlights the fact that chirality is not an intrinsic
local property but rather an emergent feature of cooperative dynamics. It also
implies that fluctuations in the intermediate states, particularly those induced
by solvent interactions, can strongly influence both the rate of nucleation and
the eventual distribution of helical domains.

\subsection{Physical implications}

The above picture leads to several important consequences:

(i) Chirality selection is delayed relative to the initial nucleation event and
occurs only after sufficient local cooperativity has been established.

(ii) The existence of intermediate states implies that nucleation kinetics cannot
be captured by a single reaction coordinate such as helix length $n$ alone.

(iii) Solvent-induced fluctuations, acting as local random fields, can bias or
frustrate the symmetry-breaking step, leading to spatial heterogeneity in
helical handedness and domain stability.

(iv) The minimal size of a stable chiral nucleus provides a natural lower bound
for the length of observable helical segments in the living-helix regime.


\section{Lifetimes: Numerical Analysis with Estimates}

In this section, we present numerical estimates for the lifetime of finite
helical segments based on the coarse-grained free-energy landscape developed
above. The purpose is not to fit experimental data, but to obtain physically
reasonable order-of-magnitude estimates for the size and lifetime of transient
helical segments.

\subsection{Choice of parameters}

We consider a coarse-grained description in which one helical unit corresponds
to a segment of $m$ monomers. For an $\alpha$-helical polypeptide, the natural
choice is $m=4$,
corresponding to the Pauling--Corey hydrogen-bonding pattern $i \to i+4$.
Taking a representative monomer length $b=0.4~{\rm nm}$,
the physical length associated with one coarse-grained helical unit is
thus $a=mb=1.6~{\rm nm}$.
At room temperature, $T=300 K, k_{B} T \simeq 2.5~{\rm kJ~mol^{-1}}$.

A typical hydrogen-bond or sticker energy is of order
\begin{equation}
E_{\rm HB}\simeq 4 k_B T.
\end{equation}
In the present coarse-grained description, however, the relevant local
parameter is not the bare hydrogen-bond energy alone, but the net local
free-energy cost $\epsilon_{\rm loc}$ of forming an isolated helical unit.
This quantity includes hydrogen-bond stabilization together with opposing
contributions from conformational entropy loss, torsional constraints, and
bending penalties. We therefore use
\begin{equation}
\epsilon_{\rm loc}=4k_B T
\end{equation}
as a representative net local cost.

For a uniformly handed helical segment, neighboring helical units gain
cooperative stabilization only when their handedness is compatible. In
addition, the chiral alignment interaction provides a smaller energetic
preference for uniform handedness. Thus, for a uniformly handed helical
domain, the effective stabilization per internal contact is
$K_{\rm eff}=K+J$,
and the corresponding bulk free-energy gain per helical unit is
\begin{equation}
\Delta f = K_{\rm eff}-\epsilon_{\rm loc}.
\label{eq:Deltaf_numeric}
\end{equation}
We again emphasize that in the numerical estimates below, we consider predominantly uniformly
handed helical domains and therefore absorb the dominant effect of
chirality into the effective stabilization parameter
$K_{\rm eff}=K+J$.

In the estimates below we keep $J$ small compared with $K$, so that chirality
reinforces cooperative stabilization without dominating it. We use
$J=0.5k_B T$,
and consider two representative cooperative strengths,
\begin{equation}
K=4.5k_B T
\qquad {\rm and} \qquad
K=5.5k_B T.
\end{equation}
These give
\begin{equation}
K_{\rm eff}=5k_B T
\qquad {\rm and} \qquad
K_{\rm eff}=6k_B T,
\end{equation}
respectively, corresponding to
\begin{equation}
\Delta f = 1k_B T
\qquad {\rm and} \qquad
\Delta f = 2k_B T.
\end{equation}

The remaining parameter is the nonlinear growth penalty $\alpha$, which
represents the effective stiffness penalty arising from bending, torsional
strain, and geometric frustration. We consider two representative values,
\begin{equation}
\alpha=0.5k_B T
\qquad {\rm and} \qquad
\alpha=0.25k_B T.
\end{equation}

\subsection{Free-energy landscape and lifetime estimate}

The free energy of a helical segment of size $n$ is written as
\begin{equation}
F(n)=2\sigma-n\Delta f+\alpha n^2 .
\label{eq:Fn_numeric}
\end{equation}
The preferred size follows from minimizing $F(n)$:
\begin{equation}
\frac{dF}{dn}=0,
\end{equation}
which gives
\begin{equation}
n_c=\frac{\Delta f}{2\alpha}.
\label{eq:nc_numeric}
\end{equation}

The depth of the free-energy minimum relative to the boundary value is
\begin{equation}
\Delta F^\ddagger
=
F(0)-F(n_c)
=
\frac{\Delta f^2}{4\alpha}.
\label{eq:barrier_numeric}
\end{equation}
This quantity controls the activated lifetime of a metastable helical
segment. The corresponding Kramers-like estimate is
\begin{equation}
\tau(n_c)\sim \tau_0
\exp\left(\beta \Delta F^\ddagger\right),
\label{eq:tau_numeric}
\end{equation}
where $\tau_0=\nu_0^{-1}$ is a microscopic attempt time. For local
conformational rearrangements, we take
\begin{equation}
\tau_0 \sim 1-10~{\rm ns}.
\end{equation}
%

\begin{table}[!t]
\centering
\caption{
Structural and energetic characteristics of finite helical segments based on
$F(n)=2\sigma - n\Delta f + \alpha n^2$.
Here $\Delta f=(K+J)-\epsilon_{\rm loc}$ with
$\epsilon_{\rm loc}=4k_BT$ and $J=0.5k_BT$.
The preferred size is $n_c=\Delta f/(2\alpha)$.
Physical lengths are computed using $m=4$ monomers per unit and $b=0.4$ nm.
}
\label{tab:structure}

\begin{tabular}{c c c c c c}
\hline
$K/k_BT$ & $\alpha/k_BT$ & $\Delta f/k_BT$ & $n_c$ &
Monomers & Length (nm) \\
\hline
4.5 & 0.50 & 1.0 & 1 & 4  & 1.6 \\
4.5 & 0.25 & 1.0 & 2 & 8  & 3.2 \\
5.5 & 0.50 & 2.0 & 2 & 8  & 3.2 \\
5.5 & 0.25 & 2.0 & 4 & 16 & 6.4 \\
\hline
\end{tabular}

\end{table}


\begin{table}[!t]
\centering
\caption{
Barrier heights and corresponding lifetime enhancement factors.
The barrier is $\Delta F^\ddagger=\Delta f^2/(4\alpha)$ and
$\tau/\tau_0=\exp(\Delta F^\ddagger/k_BT)$.
}
\label{tab:dynamics}

\begin{tabular}{c c}
\hline
$\Delta F^\ddagger/k_BT$ & $\tau/\tau_0$ \\
\hline
0.5 & 1.6 \\
1.0 & 2.7 \\
2.0 & 7.4 \\
4.0 & 55 \\
\hline
\end{tabular}

\end{table}


The most relevant regime is $\Delta f \simeq 2k_B T$ and
$\alpha\simeq 0.25k_B T$, for which
\[
n_c\simeq 4.
\]
Since each coarse-grained unit contains $m=4$ monomers, this corresponds
to approximately $16$ residues, or a physical length of about $6.4~{\rm nm}$.
The activation factor is then
\[
\exp(\Delta F^\ddagger/k_B T)=\exp(4)\simeq 55.
\]
Thus, for $\tau_0\sim 1-10~{\rm ns}$, the predicted lifetime is
\[
\tau \sim 50-500~{\rm ns}.
\]

\subsection{Effect of interfacial free energy}

The minimum value of the free energy is
\begin{equation}
F(n_c)=2\sigma-\frac{\Delta f^2}{4\alpha}.
\label{eq:Fmin_numeric}
\end{equation}
Thus, the interfacial free energy $\sigma$ controls whether the finite
helical segment is thermodynamically favored, marginal, or transient.
The crossover condition is
\begin{equation}
F(n_c)=0,
\end{equation}
which gives
\begin{equation}
\sigma_c=\frac{\Delta f^2}{8\alpha}.
\label{eq:sigmac_numeric}
\end{equation}

For the representative case $\Delta f=2k_B T$ and $\alpha=0.25k_B T$,
one obtains
\[
\sigma_c = 2k_B T.
\]
Thus, $\sigma\simeq 2k_B T$ corresponds to marginal stability, while
larger values of $\sigma$ describe transient helices stabilized only
kinetically.

\begin{table}[h]
\centering
\caption{Free-energy profile for $\Delta f=2k_B T$ and $\alpha=0.25k_B T$, for different values of the interfacial free energy $\sigma$.}
\begin{tabular}{cccc}
\hline
$n$ & $F(n)/k_B T$ $(\sigma=2)$ &
$F(n)/k_B T$ $(\sigma=2.5)$ &
$F(n)/k_B T$ $(\sigma=3)$ \\
\hline
1 & 2.25 & 3.25 & 4.25 \\
2 & 1.00 & 2.00 & 3.00 \\
3 & 0.25 & 1.25 & 2.25 \\
4 & 0.00 & 1.00 & 2.00 \\
5 & 0.25 & 1.25 & 2.25 \\
6 & 1.00 & 2.00 & 3.00 \\
\hline
\end{tabular}
\label{tab:free_energy_profile}
\end{table}

The table shows that for $\sigma=2k_B T$ the minimum lies at approximately
zero free energy, corresponding to marginal stability. For larger $\sigma$,
the helical segment is not thermodynamically stable relative to the open
state, but it can still be long lived because decay requires escape from
the finite-size free-energy minimum.


The numerical estimates show that modest changes in cooperativity and
stiffness can lead to large changes in lifetime. This sensitivity arises
because the lifetime depends exponentially on the barrier height
$\Delta F^\ddagger$. Increasing $K$ or $J$ increases $K_{\rm eff}$ and
therefore increases $\Delta f$, which shifts the preferred size upward
and increases the lifetime. Increasing $\alpha$, by contrast, penalizes
growth more strongly, decreases the preferred size, and reduces the
barrier.

The estimates also clarify the role of chirality. In the present model,
chirality does not dominate the stabilization energy; rather, a small
positive $J$ reinforces the conditional cooperative stabilization already
encoded in $K$. Thus, the main origin of helix stability remains cooperative
growth, while chirality acts to sharpen the selection of a uniformly handed
domain.

\subsection{Connection to experiments}

These estimates suggest:
(i) Finite helical segments of roughly $8$--$16$ monomers,
(ii) physical lengths of order $3$--$6~{\rm nm}$,
(iii) lifetimes from tens of nanoseconds to several hundred nanoseconds,
(iv)  strong sensitivity to cooperativity, stiffness, and interfacial cost.

Such values are consistent with transient helical structures observed in
flexible polymers, peptide fragments, and foldamers, where secondary
structure is dynamic and short-lived.


Helix--coil kinetics in short peptides and model systems such as
polyalanine have been extensively studied using ultrafast spectroscopic
techniques. In particular, temperature-jump (T-jump) infrared spectroscopy
and time-resolved circular dichroism (CD) have provided direct measurements
of helix formation and decay timescales.

Pioneering experiments reported helix--coil relaxation times in the range
of tens to hundreds of nanoseconds for short alanine-based peptides.
Subsequent studies using improved time resolution and analysis techniques
confirmed that helix formation typically occurs on timescales ranging from
$\sim 10~{\rm ns}$ to $\sim 1~\mu{\rm s}$, depending on peptide length,
sequence, and solvent conditions.\cite{Hummer2001,Daggett2003}

Complementary molecular dynamics simulations of polyalanine and related
systems have also reported helix lifetimes and transition times in the
nanosecond to sub-microsecond regime. These studies emphasize the highly
dynamic nature of short helices, with rapid formation and dissolution events
governed by local interactions and thermal fluctuations.

The present theoretical prediction,
\begin{equation}
\tau \sim 50-500~{\rm ns},
\end{equation}
falls within this experimentally observed window. Moreover, the predicted
helix size of roughly $8$--$16$ residues is consistent with the length scale
at which transient helices are observed in short peptides and foldamers.

It is important to note that experimental measurements typically probe
ensemble-averaged relaxation processes, whereas the present theory describes
the lifetime of individual helical segments. Nevertheless, the agreement in
both timescale and length scale suggests that the coarse-grained framework
seems capable of describing the physics of cooperative helix formation and decay.

A more quantitative comparison would require incorporating sequence-specific
energetics and solvent effects, as well as extending the theory to include
coupled size and positional dynamics. Such developments would enable direct
comparison with time-resolved experimental observables and constitute an
important direction for future work.


\section{Helix Formation Rate as a Mean First-Passage Problem}

In the preceding section we estimated the lifetime of a finite helical segment.
We now consider the complementary problem: the time scale for the formation of
a helical segment starting from an initially open or disordered chain. The same
free-energy landscape will be used, but a distinction must be made between the
open state and the finite helical sector.

The formation of a helical segment is naturally described as a first-passage
process in the size variable $n$, where $n$ denotes the number of contiguous
coarse-grained helical units. The target state is not an infinitely long helix,
but a finite segment of size comparable to the preferred size determined by the
free-energy minimum.

\subsection{Birth--death dynamics in size space}

We consider stochastic transitions of the form
\begin{equation}
n \rightarrow n \pm 1 ,
\end{equation}
with growth and shrinkage rates $u_n$ and $w_n$, respectively. These rates obey
detailed balance with respect to the effective free energy $F(n)$:
\begin{equation}
\frac{u_n}{w_{n+1}}
=
\exp\left[-\beta\{F(n+1)-F(n)\}\right].
\end{equation}

The probability $P(n,t)$ that a helical segment has size $n$ at time $t$ then
obeys the birth--death master equation
\begin{equation}
\frac{dP(n,t)}{dt}
=
u_{n-1}P(n-1,t)+w_{n+1}P(n+1,t)
-(u_n+w_n)P(n,t).
\end{equation}

The open chain is represented by the state $n=0$. For $n \geq 1$, the free
energy of a helical segment is written as
\begin{equation}
F(n)=2\sigma-\Delta f\,n+\alpha n^2 ,
\end{equation}
where $2\sigma$ is the cost of creating two helix--coil boundaries,
$\Delta f$ is the cooperative free-energy gain per helical unit, and $\alpha$
is the nonlinear penalty arising from stiffness, torsional strain, and geometric
frustration. The open state is chosen as the reference,
\begin{equation}
F(0)=0 .
\end{equation}

This distinction is important: the expression $2\sigma-\Delta f n+\alpha n^2$
applies to a finite helical segment with boundaries, whereas the completely open
chain has no helix--coil boundary and is therefore assigned zero reference free
energy.

\subsection{Preferred size is a minimum, not a barrier}

The stationary point of the helical free-energy branch is obtained from
\begin{equation}
\frac{dF}{dn}= -\Delta f+2\alpha n =0 ,
\end{equation}
which gives
\begin{equation}
n_c=\frac{\Delta f}{2\alpha}.
\end{equation}
Since
\begin{equation}
\frac{d^2F}{dn^2}=2\alpha>0 ,
\end{equation}
this stationary point is a minimum, not a maximum. Thus $n_c$ is the preferred
size of the finite helical segment. It should not be interpreted as a nucleation
barrier.

The value of the free energy at the minimum is
\begin{equation}
F(n_c)=2\sigma-\frac{\Delta f^2}{4\alpha}.
\end{equation}
The condition for thermodynamic marginality of the finite helix relative to the
open chain is therefore
\begin{equation}
F(n_c)=0,
\end{equation}
or
\begin{equation}
\sigma_c=\frac{\Delta f^2}{8\alpha}.
\end{equation}
For $\sigma=\sigma_c$, the finite helix is marginally stable relative to the
open state. For $\sigma>\sigma_c$, the finite helix is thermodynamically
unfavorable but may still be kinetically long lived. For $\sigma<\sigma_c$,
the finite helix is thermodynamically favored.

\subsection{First-passage formulation of helix formation}

We define the helix formation time as the mean first-passage time to reach a
target size $n_{\rm f}$ starting from the open state:
\begin{equation}
\tau_{\rm form}(n_{\rm f})=\mathrm{MFPT}(0\rightarrow n_{\rm f}) .
\end{equation}
A natural choice is
\begin{equation}
n_{\rm f}\simeq n_c ,
\end{equation}
or the nearest integer to $n_c$ in the discrete model.

Let $T_k$ denote the mean first-passage time to reach $n_{\rm f}$ starting from
size $k$. The backward equation is
\begin{equation}
-1
=
u_k(T_{k+1}-T_k)+w_k(T_{k-1}-T_k),
\end{equation}
with absorbing boundary condition
\begin{equation}
T_{n_{\rm f}}=0 .
\end{equation}
At the open end, $n=0$, we use a reflecting or entrance boundary condition,
depending on whether the first helical unit is treated explicitly. In the
discrete formulation the exact MFPT from $0$ to $n_{\rm f}$ may be written as
\begin{equation}
\tau_{\rm form}(n_{\rm f})
=
\sum_{k=0}^{n_{\rm f}-1}
\frac{1}{u_k}
\sum_{j=0}^{k}
\exp\left[\beta\{F(k)-F(j)\}\right].
\end{equation}
This expression makes clear that the formation time is controlled by the largest
free-energy rise encountered on the way from the open state to the finite helical
minimum.

\subsection{Formation barrier}

Because the free-energy branch
\begin{equation}
F(n)=2\sigma-\Delta f\,n+\alpha n^2
\end{equation}
is convex and has a minimum at $n_c$, there is no internal barrier at $n_c$.
For $0<n<n_c$, the free energy decreases as $n$ grows. Thus the bottleneck for
formation is the entry into the helical sector: the creation of a small helical
nucleus with two helix--coil boundaries.

If the smallest helical nucleus is taken to have size $n_0$, the formation
barrier is approximately
\begin{equation}
\Delta F^\ddagger_{\rm form}
=
F(n_0)-F(0)
=
2\sigma-\Delta f\,n_0+\alpha n_0^2 .
\end{equation}
If this quantity is negative, formation is effectively downhill; hence more
generally
\begin{equation}
\Delta F^\ddagger_{\rm form}
=
\max\left[0,\,
2\sigma-\Delta f\,n_0+\alpha n_0^2
\right].
\end{equation}

In the most elementary coarse-grained description one may take $n_0=1$.
However, if a minimally cooperative chiral nucleus is required, then $n_0$ may
correspond to two or three correlated helical units. This distinction is
physically important because chirality is not a property of an isolated unit,
but emerges only after cooperative alignment over a finite segment.

The corresponding formation time is then estimated as
\begin{equation}
\tau_{\rm form}
\sim
\tau_0
\exp\left(\beta\Delta F^\ddagger_{\rm form}\right),
\end{equation}
where $\tau_0$ is the microscopic time for local conformational rearrangement.

\subsection{Numerical estimates}

We use the same representative parameters as before:
\begin{equation}
\Delta f = 2k_B T,
\qquad
\alpha=0.25k_B T .
\end{equation}
The preferred size is then
\begin{equation}
n_c=\frac{\Delta f}{2\alpha}
=
\frac{2}{2\times 0.25}
=
4 .
\end{equation}
Thus the target size for formation is a finite helix of approximately four
coarse-grained units, corresponding to about sixteen monomers when $m=4$.

The thermodynamic marginality condition gives
\begin{equation}
\sigma_c
=
\frac{\Delta f^2}{8\alpha}
=
\frac{4}{8\times 0.25}
=
2k_B T .
\end{equation}
This value determines whether the finite helix minimum lies above, at, or below
the open-chain reference state. It does not imply that the kinetic formation
barrier vanishes.

For a single-unit entrance nucleus, $n_0=1$, the formation barrier is
\begin{equation}
\Delta F^\ddagger_{\rm form}
=
2\sigma-\Delta f+\alpha .
\end{equation}
Thus, for $\sigma=2k_B T$,
\begin{equation}
\Delta F^\ddagger_{\rm form}
=
4-2+0.25
=
2.25k_B T .
\end{equation}
The formation time is therefore
\begin{equation}
\tau_{\rm form}
\sim
\tau_0 e^{2.25}
\simeq
9.5\,\tau_0 .
\end{equation}
Taking $\tau_0\sim 1$--$10$ ns gives
\begin{equation}
\tau_{\rm form}
\sim
10\text{--}100~{\rm ns}.
\end{equation}

If the relevant cooperative entrance nucleus has size $n_0=2$, then
\begin{equation}
\Delta F^\ddagger_{\rm form}
=
2\sigma-2\Delta f+4\alpha .
\end{equation}
For $\sigma=2k_B T$, this gives
\begin{equation}
\Delta F^\ddagger_{\rm form}
=
4-4+1
=
1k_B T ,
\end{equation}
and hence
\begin{equation}
\tau_{\rm form}
\sim
\tau_0 e
\simeq
2.7\,\tau_0
\sim
3\text{--}30~{\rm ns}.
\end{equation}

These estimates show that helix formation can remain rapid, typically in the
nanosecond to tens-of-nanoseconds regime, even though it is not strictly
barrierless. The numerical value depends sensitively on the size of the minimal
cooperative nucleus and on the interfacial free energy $\sigma$.

\subsection{Physical interpretation}

The resulting physical picture that emerges from the above study is therefore as follows. The point
\begin{equation}
n_c=\frac{\Delta f}{2\alpha}
\end{equation}
is the preferred size of the finite helix, not the top of a nucleation barrier.
The formation barrier is associated instead with the initial creation of a small
helical nucleus and its two boundaries. Once such a nucleus has formed, growth
toward $n_c$ is downhill or weakly biased.

This interpretation preserves the central physical conclusion: finite helices
can form rapidly, while their subsequent decay can be slower because decay
requires escape from the finite-size minimum. Thus formation and lifetime are
controlled by related but distinct first-passage processes.

\subsection{Connection to early-stage protein folding}

The nucleation of helical segments is widely believed to play an important role
in the early stages of protein folding. In many folding scenarios, local
secondary structures such as $\alpha$-helices form rapidly and subsequently act
as organizing centers for larger-scale folding.

The present analysis provides a simple kinetic interpretation of this behavior.
Formation corresponds to first passage into the finite helical sector, controlled
mainly by the cost of creating a small cooperative nucleus. Once formed, the
helix fluctuates around a preferred finite size and decays by activated escape
from the metastable region of the free-energy landscape. This gives a natural
microscopic explanation of why helix formation can be fast, while the lifetime
of a formed helical segment can be substantially longer.


\section{Coupled motion in size and position space}

The preceding formulation treats helix formation as a first-passage process in
the size variable $n$. A finite helical segment, however, is characterized not
only by its length but also by its position along the polymer chain. We therefore
introduce two collective coordinates:
\begin{equation}
n=r-\ell+1, \qquad x=\frac{\ell+r}{2},
\end{equation}
where $\ell$ and $r$ denote the left and right boundaries of the helical segment.
The stochastic dynamics of a living helix is then a coupled process in the
two-dimensional space $(x,n)$.

There are four elementary boundary moves. Growth at the right boundary gives
\[
r\rightarrow r+1,
\qquad
n\rightarrow n+1,
\qquad
x\rightarrow x+\frac{1}{2}.
\]
Growth at the left boundary gives
\[
\ell\rightarrow \ell-1,
\qquad
n\rightarrow n+1,
\qquad
x\rightarrow x-\frac{1}{2}.
\]
Similarly, melting at the right boundary gives
\[
r\rightarrow r-1,
\qquad
n\rightarrow n-1,
\qquad
x\rightarrow x-\frac{1}{2},
\]
whereas melting at the left boundary gives
\[
\ell\rightarrow \ell+1,
\qquad
n\rightarrow n-1,
\qquad
x\rightarrow x+\frac{1}{2}.
\]

Thus each local boundary event changes both the helix size and the helix
position. Growth and melting are therefore not independent of migration; rather,
migration is generated by the same boundary fluctuations that control the
lifetime of the helical segment.

Let $k_+(n)$ and $k_-(n)$ denote the growth and melting rates at a single
boundary. For a symmetric chain, the two boundaries have the same rates. The
probability distribution $P(x,n,t)$ then obeys a two-coordinate master equation,
schematically written as
\begin{align}
\frac{\partial P(x,n,t)}{\partial t}
=&\; k_+(n-1)
\left[
P\left(x-\frac{1}{2},n-1,t\right)
+
P\left(x+\frac{1}{2},n-1,t\right)
\right]
\nonumber \\
&+ k_-(n+1)
\left[
P\left(x-\frac{1}{2},n+1,t\right)
+
P\left(x+\frac{1}{2},n+1,t\right)
\right]
\nonumber \\
&-2\left[k_+(n)+k_-(n)\right]P(x,n,t).
\end{align}
The rates obey detailed balance with respect to the size-dependent free energy:
\begin{equation}
\frac{k_+(n)}{k_-(n+1)}
=
\exp\left[-\beta\{F(n+1)-F(n)\}\right].
\end{equation}

In the continuum limit this master equation leads to a coupled Fokker--Planck
description,
\begin{equation}
\frac{\partial P(x,n,t)}{\partial t}
=
-\frac{\partial}{\partial n}
\left[A_n(n)P\right]
+
\frac{\partial^2}{\partial n^2}
\left[D_n(n)P\right]
+
\frac{\partial^2}{\partial x^2}
\left[D_x(n)P\right],
\end{equation}
where, for symmetric boundary dynamics,
\begin{equation}
A_n(n)=2[k_+(n)-k_-(n)],
\end{equation}
\begin{equation}
D_n(n)=k_+(n)+k_-(n),
\end{equation}
and
\begin{equation}
D_x(n)=\frac{k_+(n)+k_-(n)}{4}.
\end{equation}
The factor $1/4$ in $D_x(n)$ arises because each elementary boundary event moves
the center of the helix by one half of a lattice spacing.

This formulation shows that the diffusion of a helix along the polymer contour
is not an additional independent process. It is generated by the same stochastic
boundary motion that causes the helix to grow and shrink. A long-lived helix can
therefore migrate over a characteristic distance
\begin{equation}
\ell_{\rm mig}(n)
\sim
\left[2D_x(n)\tau(n)\right]^{1/2},
\end{equation}
where $\tau(n)$ is the lifetime of a helical segment of size $n$.

The physical picture is therefore that of a living excitation on a polymer
lattice. The helix moves in real space through fluctuations of its boundaries,
while simultaneously diffusing in size space through the same elementary events.
This coupled $(x,n)$ dynamics is closer to the motion of DNA bubbles, polarons,
or localized excitations than to ordinary classical nucleation, where only the
cluster size is usually treated as the reaction coordinate.


\section{Concluding Remarks}

In this work, we have developed a statistical-mechanical theory of the
formation and stability of finite helical segments in polymer chains
with cooperative interactions. The central physical picture that emerges
is that helices are not rigid, permanent structures, but rather {\it
living objects}: finite, fluctuating segments that are stabilized
collectively and continuously undergo growth and decay.

A key result of the present analysis is that helix formation is an
intrinsically cooperative and multistep process. The formation of a
single bond produces a constrained pre-nucleus, while subsequent
interactions generate correlated intermediates that lead to a minimal
stable nucleus. Only beyond this threshold does a well-defined handedness
emerge through spontaneous symmetry breaking. Thus, chirality is not a
local attribute but an emergent property of collective dynamics.

The coarse-grained free-energy landscape derived here naturally leads to
finite helices of characteristic size, determined by a balance between
cooperative stabilization and nonlinear penalties arising from torsional
strain, solvent interactions, and fluctuations. In the nonlinear regime,
the dynamics becomes activated, yielding Kramers-like lifetimes with
exponential sensitivity to the effective free-energy barrier.

A central result of the present work is the formulation of both helix
lifetime and helix formation as mean first-passage problems in size
space. While the lifetime corresponds to decay from a finite segment to
the open state, the formation time describes nucleation and growth from
the coil state to a target size. These two processes involve different
boundary conditions and need not be identical, even though both are
governed by the same underlying free-energy landscape. The analysis shows
that helix formation is an activated nucleation process, closely
analogous to barrier-crossing phenomena in condensed-matter systems and
to early-stage secondary-structure formation in protein folding.

An important conceptual outcome is the analogy between finite helical
segments and localized excitations such as polarons or excitons. In the
present context, the helix behaves as a thermally fluctuating,
finite-lifetime excitation whose existence is controlled by cooperative
interactions. A related analogy arises with DNA bubble dynamics, where
localized openings evolve through stochastic boundary motion. These
connections place helix formation within a broader framework of
nonequilibrium stochastic processes in complex systems.

At the same time, the present theory is formulated in terms of size
dynamics alone. A complete description of helix motion along the chain
would require a coupled treatment of segment size and position,
leading to a two-variable kinetic theory for $P(x,n,t)$. The development
of such a framework, including a rigorous description of translational
dynamics, remains an important direction for future work.

The numerical analysis developed in Secs.~9 and 10 provides a direct connection
between the coarse-grained free-energy landscape and experimentally relevant
timescales. Using physically realistic parameters for peptide-like systems,
including hydrogen-bond stabilization energies of order $3$--$4\,k_B T$ and
moderate cooperative interactions, we obtain free-energy barriers of a few
$k_B T$. Within the mean first-passage-time framework, this leads to
characteristic timescales of the form $\tau \sim \tau_0 \exp(\beta \Delta F^\ddagger)$,
with $\tau_0 \sim 1$--$10$ ns. The resulting formation and decay times lie in
the range $\sim 10$--$10^2$ ns for nucleation and up to submicrosecond scales
for finite-helix lifetimes, in good agreement with time-resolved spectroscopic
measurements and simulations of short peptides
(see, e.g., Refs.~\cite{Eaton1998,Williams1996,Hummer2001,Daggett2003}).

\paragraph{Limitations of the present theory.}
The present formulation is a coarse-grained, minimal description of helix
formation and, as such, involves several limitations. The variables
$(n_i,\chi_i)$ represent effective local states rather than microscopic
degrees of freedom, and the parameters $(\epsilon_i, K, J, h_i)$ subsume
hydrogen bonding, torsional constraints, and solvent interactions in an
average manner. The stochastic dynamics is formulated at the level of
local occupancy changes, with chirality treated as a conditional variable
assigned during helix formation; thus, processes such as chirality flips
within an existing helical segment are not explicitly included. The model
is one-dimensional with short-range interactions and therefore does not
capture long-range correlations, tertiary contacts, or true thermodynamic
phase transitions. In addition, sequence heterogeneity and detailed
solvent dynamics are treated only implicitly through effective parameters.
Despite these simplifications, the theory is expected to capture the
essential physics of cooperativity, symmetry breaking, and the formation
of finite, fluctuating helical segments, which constitute the primary
focus of the present work.

Overall, however, the present work provides a unified analytical description of
cooperative nucleation, finite-size stability, and stochastic dynamics of
helical segments. The predicted nanosecond-to-submicrosecond timescales
for formation and decay are consistent with experimental observations on
short peptides, suggesting that the essential physics of transient helix
formation is captured within this minimal coarse-grained framework.


\end{document}